\documentclass[fleqn,10pt]{wlscirep}
\usepackage[utf8]{inputenc}
\usepackage[T1]{fontenc}
\usepackage{xspace}
\usepackage{float}
\usepackage{xspace}
\usepackage{graphicx}
	\graphicspath{{./figures/}}
\usepackage{dcolumn}
\usepackage{bm}
\usepackage{color}
\usepackage{soul}
\usepackage{footnote}
\usepackage{multirow}
\usepackage{hyperref}
\hypersetup{
    unicode=false,          
    pdftoolbar=true,        
    pdfmenubar=true,        
    pdffitwindow=false,     
    pdfstartview={FitH},    
    pdftitle={Spontaneous phonon frequency combs from an isolated Einstein mode in InSiTe3},    
    pdfnewwindow=true,      
    colorlinks=true,       
    linkcolor=blue,          
    citecolor=blue,        
    filecolor=blue,      
    urlcolor=blue,       
    plainpages=false,        
}

\usepackage[english]{babel}
\usepackage{booktabs}
\usepackage{threeparttable}

\usepackage{xcolor}
\usepackage{soul}

\setstcolor{red}          
\setul{0.4ex}{1.2pt}

\newcommand{\IST}{\ensuremath{\mathrm{InSi}\mathrm{Te}_{3}}\xspace}
\newcommand{\grd}{$^{\circ}$\xspace}
\newcommand{\vei}{\ensuremath{{\mathbf e}_{i}}\xspace}
\newcommand{\ves}{\ensuremath{{\mathbf e}_{s}}\xspace}
\newcommand{\wn}{\ensuremath{\rm cm^{-1}}\xspace}
\newcommand{\Algp}{\texorpdfstring{\ensuremath{A^{(3^\prime)}_{1g}}\xspace}{A1g3'}}
\newcommand{\Algpp}{\texorpdfstring{\ensuremath{A^{(3^{\prime\prime})}_{1g}}\xspace}{A1g3''}}
\newcommand{\Alg}{\texorpdfstring{\ensuremath{A_{1g}}\xspace}{A1g}}

\newcommand{\AZg}{\texorpdfstring{\ensuremath{A_{2g}}\xspace}{A2g}}

\newcommand{\Eg}{\texorpdfstring{\ensuremath{E_{g}}\xspace}{Eg}}

\newcommand{\Alu}{\texorpdfstring{\ensuremath{A_{1u}}\xspace}{A1u}}
\newcommand{\AZu}{\texorpdfstring{\ensuremath{A_{2u}}\xspace}{A2u}}
\newcommand{\Eu}{\texorpdfstring{\ensuremath{E_{u}}\xspace}{Eu}}

\title{Phonon frequency comb close to an isolated Einstein mode in \textbf{\IST}}

\author[1]{Tea Belojica}
\author[1]{Jovan Blagojevi\'{c}}
\author[1,2]{Sanja Djurdji\'{c} Mijin}
\author[1]{Andrijana \v{S}olaji\'{c}}
\author[1]{Jelena Pe\v{s}i\'{c}}
\author[1]{Emil S. Bozin}
\author[1,3]{Bojana Vi\v{s}i\'{c}}
\author[4,5]{Yu Liu}
\author[4,6,7,8]{Cedomir Petrovic}
\author[9]{Zoran V. Popovi\'{c}}
\author[10,11]{Rudi Hackl}
\author[1,*]{Ana Milosavljevi\'{c}}
\author[1]{Nenad Lazarevi\'{c}}

\affil[1]{Center for Solid State Physics and New Materials, Institute of Physics Belgrade, Pregrevica 118, 11080 Belgrade, Serbia}
\affil[2]{Departamento de F\'{i}sica de Materiales, Facultad de Ciencias, Universidad Aut\'{o}noma de Madrid, 28049 Madrid, Spain} 
\affil[3]{Department of Condensed Matter Physics, Jozef Stefan Institute, Jamova cesta 39, Ljubljana 1000, Slovenia}
\affil[4]{Shanghai Advanced Research in Physical Sciences (SHARPS), Shanghai 201203, China}
\affil[5]{Center for Correlated Matter and School of Physics, Zhejiang University, Hangzhou 310058, People's Republic of China}
\affil[6]{Center for High Pressure Science \& Technology Advanced Research (HPSTAR) - Beijing 100094, China}
\affil[7]{Condensed Matter Physics and Materials Science Department, Brookhaven National Laboratory, Upton, NY 11973-5000, USA}
\affil[8]{Department of Nuclear and Plasma Physics, Vinca Institute of Nuclear Sciences, University of Belgrade, Belgrade 11001, Serbia}
\affil[9]{Serbian Academy of Sciences and Arts, Kneza Mihaila 35, 11000 Belgrade, Serbia}
\affil[10]{School of Natural Sciences, Department of Physics E51, Technische Universit\"{a}t M\"{u}nchen, 85748 Garching, Germany}
\affil[11]{IFW Dresden, Helmholtzstrasse 20, 01069 Dresden, Germany}
\affil[*]{ana.milosavljevic@ipb.ac.rs}

\begin{abstract}
The  emergence of  phonon frequency combs exemplifies a rare and intriguing phenomenon in quantum solids. Materials with distinctive phonon band structures are especially promising for hosting such states, as their vibrational dispersion landscape across the Brillouin zone can facilitate the formation of long-lived, collective lattice excitations. In the layered Van der Waals  compound \IST, polarization-resolved Raman spectroscopy reveals a pronounced anharmonicity in symmetry-predicted modes and the formation of a self-organized  frequency domain structure (coherent-like state), in the range of a localized high-energy \Alg phonon mode near 500 \wn. This strong phonon-phonon coupling manifests itself as an anomalous temperature dependence around 200 K, coinciding with the appearance of higher-order excitations within the phonon density of states gap. These findings position \IST as an unconventional platform where intrinsic highly structured phonon spectral correlations and unusually strong anharmonic effects coexist, opening new avenues for exploring emergent vibrational phenomena in low-dimensional materials.
\end{abstract}

\begin{document}
\flushbottom
\maketitle
\thispagestyle{empty}

\section*{Introduction}

Femtosecond pump–probe spectroscopy has long been the primary tool for generating and studying coherent phonons. In these experiments, an ultrashort laser pulse  excites the lattice, initiating collective vibrational motion that can be tracked in real time.\cite{PhysRevB.45.768, nano10122543, Misochko2001, PhysRevB.108.224309}When a femtosecond laser pulse reshapes the electronic potential, the equilibrium atomic positions are displaced, leaving the atoms to oscillate coherently around the new minimum. This displacive excitation mechanism inherently favors fully symmetric modes. As a result, materials like bismuth and antimony consistently exhibit strong \Alg coherent phonons in pump–probe experiments, while lower-symmetry modes appear only under special conditions.\cite{10.1063/1.2940130, 10.1063/1.2363746, PhysRevLett.77.3661} In contrast, finding equidistant closely separated phonon lines or a phonon comb  without femtosecond excitation is unusual and reflects a nonlinear lattice potential.

Layered Van der Waals (VdW) materials, with their quasi–low-dimensional character, provide an important platform for studying emergent lattice, electronic, and magnetic phenomena.\cite{Lazarević_2020, Yu_Liu_magnetic_VdW_acsnano} The class is broad, from transition metal trihalides (CrI$_3$, VI$_3$),\cite{PhysRevB.97.014420, Huang2017, PhysRevB.101.024411, SanjaVI3, PhysRevB.98.104307} transition metal trichalcogenides (CrSiTe$_3$, CrGeTe$_3$),\cite{OUVRARD198827, Zhang_2016, PhysRevB.98.104306, CSGT} iron-based tellurides (Fe$_3$GeTe$_2$) \cite{Fei2018} to transition metal dichalcogenides (VSe$_2$, MnSe$_2$).\cite{Bonilla2018, OHara2018} The acceleration in the discovery of such quasi-2D systems provides many opportunities to study  phonon correlations and nonlinear phenomena as an intrinsic property of the lattice and at the same time to consider their potential for future applications.\cite{Sierra2021, Yang_vdW, AppElemental2D, C9NR05919A} During the course of this research, Chen \textit{et al.} reported the spontaneous formation of phonon frequency combs in CrSiTe$_3$ and CrGeTe$_3$,\cite{Chen2025} emphasizing the growing interest in the vibrational properties of ternary VdW trichalcogenides. These developments further emphasize the relevance of exploring related compounds such as InSiTe$_3$, where nonlinearities of the lattice potential  may entail unconventional vibrational dynamics.

The first report of \IST single-crystal synthesis appeared more than three decades ago.\cite{IST_Sandre} Since then, its related compounds such as CrSiTe$_3$, CrGeTe$_3$, InGeTe$_3$, and AlSiTe$_3$ have been extensively studied, both theoretically and experimentally, for their  magnetic, electronic, and vibrational properties.\cite{Korkmaz2021, C7TA04810F, SANDRE1994145, Casto2015_AM3_041515, PhysRevLett.127.217203, PhysRevB.101.205125} By contrast, only a few studies have addressed the fundamental properties of \IST.\cite{C7TA04810F, Korkmaz2021, SURIWONG201875, PhysRevB.93.245307} Nevertheless, it has already been applied in broadband photodetectors with ultrafast response times \cite{doi:10.1021/acsnano.1c11628} which makes a deeper understanding of its lattice dynamics particularly relevant. Low-temperature experiments have not been performed on \IST, leaving certain aspects of the physical phenomena below room temperature unexplored.

In this work, we investigate the lattice dynamics of the VdW compound InSiTe$_3$ using temperature-dependent polarization-resolved Raman scattering combined with density functional theory (DFT) calculations. Our study reveals a pronounced anharmonicity and the formation of frequency-domain phonon comb associated with a localized high-energy \Alg mode near 500 cm$^{-1}$. This state reflects a frequency-locked spectral response of the SiTe$_3$ tetrahedra, linked to a flat and isolated phonon branch. The temperature-dependent analysis further uncovers anomalous linewidth behavior around 200~\,K, together with large values of \Alg phonon coupling constants. Upon heating to this temperature, broad features emerge in the parallel scattering configuration within the gap of the calculated phonon density of states (PDOS), consistent with overtone excitations. These findings identify InSiTe$_3$ as an important member of the of VdW family and underscore the role of   anharmonic interactions in VdW trichalhogenides.

\section*{Results}

\begin{figure}[b!]
  \centering
  \includegraphics[width=85mm]{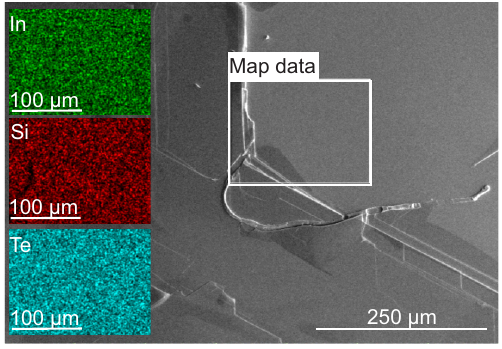}
  \caption{SEM and EDS mapping of a freshly cleaved surface of an \IST single crystal. The right part of the figure shows a flat surface over an extended area. The white rectangle indicates the area in which the EDS mapping was performed. The green, red, and turquoise areas on the left demonstrate the homogeneous distributions of the elements.}
 \label{fig:Figure1}
\end{figure}

The Raman measurements were performed on freshly cleaved surfaces. As shown in Fig.~\ref{fig:Figure1} scanning electron microscopy (SEM) reveals a flat surface of at least $0.5\times0.5~{\rm mm}^2$. The individual maps obtained with energy dispersive spectroscopy (EDS) as displayed on the left hand side of Fig.~\ref{fig:Figure1} demonstrate the uniformity of the sample over several tens of microns and corroborate the atomic ratio of In:Si:Te to be 1:1:3. No impurities,  contaminations and vacancies could be detected.

For the phonon analysis, the symmetry positions of the atoms in the lattice are relevant. The factor group analysis for the $P\overline{3}1m$ space group (No. 162)\cite{Jin2018} yields the symmetry allowed phonons in the center of the Brillouin zone (BZ). The corresponding Raman tensors that govern selection rules are listed in Tab.~\ref{ref:Table1}. 
In our experiment, the plane of incidence is the \(ab\)-plane, where \(|a| = |b|\) and \(\measuredangle(a,b) = 120^\circ\). This plane is oriented perpendicular to the optical axis of the spectrometer, so that the light propagates along the \(c\)-axis. The incident and scattered light polarizations, \(\mathbf{e}_\mathrm{i}\) and \(\mathbf{e}_\mathrm{s}\), respectively, lie in the \(ab\)-plane, with \(\mathbf{e}_\mathrm{i} \parallel \mathbf{e}_\mathrm{s}\) corresponding to \(\theta = 0^\circ\) and \(\mathbf{e}_\mathrm{i} \perp \mathbf{e}_\mathrm{s}\) corresponding to \(\theta = 90^\circ\). In these two scattering configurations all allowed Raman active modes ($3\Alg + 5\Eg$) can be observed. The doubly degenerate \Eg modes are observable in both parallel and crossed polarizations, whereas the fully symmetric \Alg modes vanish in configurations with crossed polarization.
In addition to the experiments, density-functional theory (DFT) calculations were performed. The resulting frequencies of the optical phonons are in good agreement with the experimental values obtained at 80 K, as summarized in Tab.~\ref{ref:Table2}.
\begin{table*}[t!]
    \centering
    \renewcommand{\arraystretch}{1.3}
    \setlength{\tabcolsep}{20pt}
    \caption{Atoms, Wyckoff positions, related $\Gamma$ point phonons, and Raman tensors for the $P\overline{3}1m$ space group.}
    \label{ref:Table1}
    \begin{threeparttable}
        \begin{tabular}{c c c}
            \toprule
            \multicolumn{1}{c} {Atoms (Wyckoff positions)} & \multicolumn{2}{c} {Irreducible representations} \\
            \midrule
            \multicolumn{1}{c}{In ($2d$)} & \multicolumn{2}{c}{$\AZu + \Eu + \Eg$} \\[1mm]
			\multicolumn{1}{c}{Si ($2e$)} & \multicolumn{2}{c}{$\AZu + \Eu + \Alg + \Eg$}\\[1mm]
			\multicolumn{1}{c}{Te ($6k$)} & \multicolumn{2}{c}{$2\AZu + 3\Eu + 2\Alg + 3\Eg$}\\[1mm]
			\midrule
			\multicolumn{3}{c}{Raman tensors} \\[1mm]
			\midrule
			\multicolumn{3}{c}{$
			A_{1g} = \begin{pmatrix}
			a&0&0\\
			0&a&0\\
			0&0&b\\
			\end{pmatrix}
			$} \\
			$
			^1E_{g} = \begin{pmatrix}
			c&0&0\\
			0&-c&d\\
			0&d&0\\
			\end{pmatrix}
			$
			&
			&
			$
			^2E_{g} = \begin{pmatrix}
			0&-c&-d\\
			-c&0&0\\
			-d&0&0\\
			\end{pmatrix}$\\
			 \bottomrule
        \end{tabular}
    \end{threeparttable}
\end{table*}

\begin{figure}[t!]
  \centering
  \includegraphics[width=85mm]{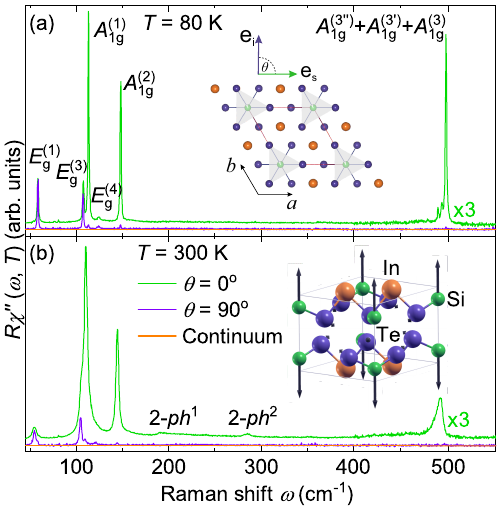}
  \caption{Raman spectra of \IST in parallel ($\theta = 0^{\circ}$) and cross ($\theta = 90^{\circ}$) polarization configurations at (a) 80\,K and (b) 300\,K. The orange lines represent the phenomenological continua (see text). Inset of (a) \IST crystallographic unit cell with vectors of incident and scattered light polarizations \vei and \ves, respectively. For symmetry reasons the orientation of the polarizations with respect to the crystal axes $a$ and $b$  is irrelevant. Inset of (b) displacement pattern of $\Alg^{(3)}$ mode. The arrow lengths are proportional to the square root of the inter-atomic forces. For this mode only the Si atoms move.}
 \label{fig:Figure2}
\end{figure}

\begin{table*}[t!]
    \centering
    \renewcommand{\arraystretch}{1.3}
    \setlength{\tabcolsep}{20pt} 
    \caption{Phonon symmetry, activity, experimental (80\,K) and theoretical phonon frequencies (0\,K) calculated using experimental crystallographic data. All values are given in \wn. Silent modes are indicated by an asterisk.}
    \label{ref:Table2}
    \begin{threeparttable}
        \begin{tabular}{c c c c c}
            \toprule
            \multicolumn{5}{c} {Space group:  $P\overline{3}1m$ (No. 162)} \\
            \midrule
            \multicolumn{3}{c} {Even ($g$)} & \multicolumn{2}{c} {Odd ($u$)}\\
            \midrule
            Symmetry & Exp. & Calc. & Symmetry & Calc.\\[1mm]
            \midrule
            $\AZg^{(1)*}$ & - & 25.5 & $\AZu^{(1)}$ & 0\\[1mm]
			$\Eg^{(1)}$ & 58.5 & 57.7 & $\Eu^1$ & 0\\[1mm]
			$\Eg^{(2)}$ & - & 77.0 & $\AZu^{(2)}$ & 67.1 \\[1mm]
			$\Eg^{(3)}$& 107.8 & 100.0 & $\Eu^{(2)}$ & 69.1\\[1mm]
			$\Alg^{(1)}$ & 113.4 & 108.4 & $\Alu^{(1)*}$ & 89.9\\[1mm]
			$\AZg^{(2)*}$& - & 118.2 & $\Eu^{(3)}$ & 92.1\\[1mm]
			$\Eg^{(4)}$ & 124.9 & 123.8 & $\Eu^{(4)}$ & 111.9\\[1mm]
			$\Alg^{(2)}$ & 148.3 & 145.5 & $\AZu^{(3)}$ & 154.1\\[1mm]
			$\Eg^{(5)}$& - & 360.4 & $\AZu^{(4)}$ & 239.4\\[1mm]
			$\Alg^{(3)}$ & 498.0 & 486.4 & $\Eu^{(5)}$ & 360.1\\[1mm]
			 \bottomrule
        \end{tabular}

    \end{threeparttable}
\end{table*}
Representative Raman spectra of \IST at 80~\,K and 300~\,K  obtained for both parallel ($\theta = 0$\grd) and  crossed ($\theta = 90$\grd) polarization configurations are shown in Fig.~\ref{fig:Figure2}~(a) and (b), respectively. All three \Alg modes and three out of the five expected \Eg modes appear as narrow lines. At 80~K the Raman spectrum in parallel scattering configuration [Fig.~\ref{fig:Figure2}~(a)] reveals two extra  peaks in the energy range 450 to 500~\wn labeled as \Algpp and \Algp.
Although weak in intensity, they are still distinctly observable on the low-energy side of the Raman-active $\Alg^{(3)}$ mode. These additional two peaks and the $\Alg^{(3)}$ phonon are equidistant and persist up to high temperatures, although they cannot be resolved any further as separate lines [Fig.~\ref{fig:Figure2}~(b) and Fig.~\ref{fig:Figure6}]. As seen in Fig.~\ref{fig:Figure2}(b), at 
300 K additional excitations are observed in the parallel scattering configuration, appearing slightly
below 200~\wn and 300~\wn.
\begin{figure}[b!]
  \centering
  \includegraphics[width=85mm]{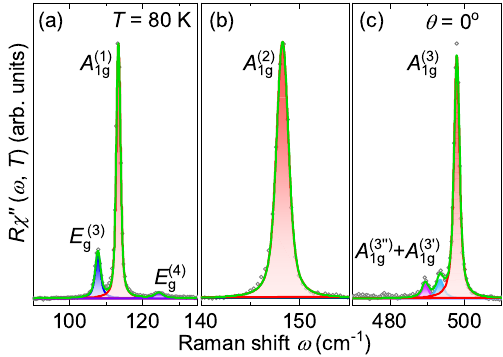}
  \caption{(a)-(c) Phonon excitations modeled with Voigt profiles in parallel ($\theta = 0^{\circ}$) polarization configuration where phonons of both \Alg and \Eg symmetry are observed. The spectra are recorded at 80\,K.}
 \label{fig:Figure3}
\end{figure}

For the quantitative analysis, the phonon peaks were modeled using a Voigt line shape, defined as the convolution of a Lorentzian and a Gaussian function. The Gaussian component, which accounts for the instrumental resolution of the spectrometer, was fixed to a linewidth of ${\Gamma}_{\mathrm{G}}=1$~\wn. To this end, the continuum was modeled phenomenologically using a Drude function together with a linear term: $\chi^{\prime\prime}_{\rm continuum}(\omega)  \propto A \Gamma \omega / (\omega^2 + {\Gamma}^2) + B \omega$ (orange lines in  Fig.~\ref{fig:Figure2}) and  was subtracted from the raw data, where $A$, $B$, and $\Gamma$ are temperature dependent phenomenological parameters. The resulting continuum is very weak and essentially constant within the phonon energy range. Fig.~\ref{fig:Figure3} displays the results of the phonon analysis in parallel scattering configuration at 80\,K. 

\begin{figure}[t!]
  \centering
  \includegraphics[width=85mm]{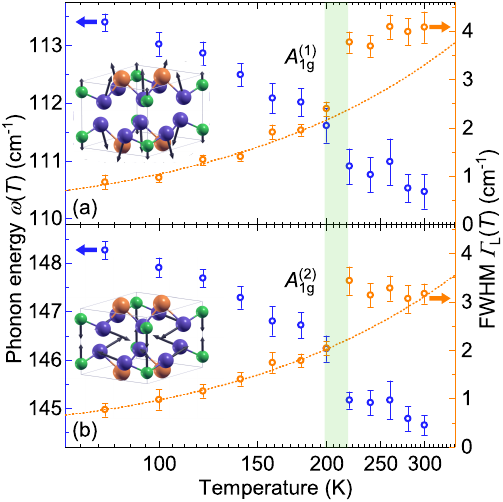}
  \caption{Temperature dependences of the energies and Lorentzian linewidths of the $\Alg^{(1)}$ and $\Alg^{(2)}$ phonons. There are discontinuities of both the energies and linewidths close to 200~K. The dashed lines represent fits to the data below 200~K. The linewidths and  energies are well described by anharmonic phonon decay  (Eq.~\ref{eq:Anharmonic}) and thermal expansion, respectively.
  }
 \label{fig:Figure4}
\end{figure}

We first focus on the low-lying \Alg modes, labeled $\Alg^{(1)}$ and $\Alg^{(2)}$. The temperature dependencies of their energies and linewidths are displayed in Fig.~\ref{fig:Figure4}, with the vibration patterns shown in insets. For describing the linewidths, we use the Klemens model of symmetric anharmonic decay, \cite{PhysRev.148.845}  

\begin{equation}
\mathrm{\Gamma_L}(T) = \mathrm{\Gamma_L(0)}\left(1+\frac{2\mathrm{\lambda_{ph-ph}}}{\mathrm{e}^{\frac{\hbar {\omega}_{0}}{2\mathrm{k_B}T}}-1}\right),
\label{eq:Anharmonic}
\end{equation}

\noindent where $\mathrm{\Gamma_L(0)}$ and ${\omega}_{0}$ were obtained by extrapolating linewidths and energies to the zero temperature limit. The phonon-phonon coupling parameter $\mathrm{\lambda_{ph-ph}}$ parametrizes the interaction between the optical (here Raman-active) phonon and the acoustical modes at $\pm{\bf k}$ and $\hbar\omega/2$. The interaction may be mediated by electrons, spins or fluctuations.

At low temperatures $\Alg^{(1)}$ and $\Alg^{(2)}$ exhibit linewidths  in the range of $\Gamma_L < 1$~\wn providing further evidence of the excellent crystal quality. Across the measured temperature range, the linewidths of $\Alg^{(1)}$ and $\Alg^{(2)}$ increase nearly by a factor of four. Up to 200~K the linewidths are well described by the symmetric anharmonic decay as described by Klemens \cite{PhysRev.148.845} and described in Eq.~\ref{eq:Anharmonic}. The temperature dependence indicates strong phonon–phonon coupling beyond 1, $\mathrm{\lambda_{ph-ph}}\left(\Alg^{(1)}\right) = 1.6$ and $\mathrm{\lambda_{ph-ph}}\left(\Alg^{(2)}\right) = 1.7$.\cite{PhysRevB.97.054306} In both cases, the next data points at 220~K is offset by more than a wavenumber, well beyond the statistical uncertainty. Up to room temperature the linewidths are essentially temperature independent. A similar (and consistent)  discontinuity is observed for the energies.  The cause of these discontinuities is presently unresolved, with no phase transition reported in this temperature range to the best of our knowledge.
\begin{figure}[ht]
  \centering
  \includegraphics[width=85mm]{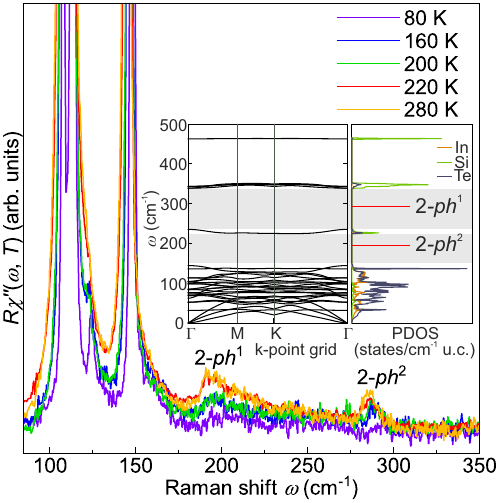}
  \caption{Raman spectra in the range between 80~\wn and 350~\wn at temperatures as indicated. The overtone excitations increase abruptly between 200 and 220~K in intensity. Inset: Calculated phonon dispersion along the high-symmetry directions as indicated and PDOS. The shaded area marks the gap in the PDOS.}
 \label{fig:Figure5}
\end{figure}

Even so, the Raman spectra in the range between 80~\wn and 350~\wn exhibit an intensity anomaly in the same temperature range as shown in Fig.~\ref{fig:Figure5}. New features at 200~\wn and 280~\wn emerge with increasing temperature. Between 80 and 200~K these broad structures gain intensity slowly and gradually, while they double in intensity between 200 and 220~K and then remain constant up to 300~K. Upon inspecting the phonon dispersion and density of states (PDOS) (see inset of Fig.~\ref{fig:Figure5}) we conclude that these bands derive from the relatively flat bands in the range of 100~\wn and the nearly dispersionless band slightly below 150~\wn, mainly deriving from Te vibrations, and may be identified as 2-phonon excitations in the gaps of the phonon bands. Since these 2-phonon bands derive from phonons with energies on the scale of the thermal energy ($140~\wn \approx 200$~K), one may speculate that thermally excited quasiparticles mediate an enhanced coupling between the low-energy Te modes. Then, the $\Alg^{(1)}$ and $\Alg^{(2)}$ Raman modes which just probe this increased coupling get more damped, and the linewidths increase while the energies decrease. Similar as for the Raman-active phonons, the coupling of all other modes increases across the entire Brillouin zone and boosts the two-phonon DOS. This electron-mediated coupling scenario would be consistent with the absence of any signature in the structure.

The three equidistant lines in the range of the $\Alg^{(3)}$ mode predicted by DFT are fitted with Voigt profiles and shown in Fig.~\ref{fig:Figure3}. In Fig.~\ref{fig:Figure6} they are presented on an expanded energy scale for more temperatures, both along with Voigt profiles and the frequency combs model to be described below. DFT calculations suggest that the $\Alg^{(3)}$ mode is an Einstein phonon with only Si atoms involved far above all other lines (see inset of Fig.~\ref{fig:Figure5}). However, instead of an isolated line we observe three equidistant lines in the energy range of the $\Alg^{(3)}$ mode. All three $\Alg^{(3)}$ peaks move to lower energy with increasing temperature. While the distances between the modes are essentially temperature independent, the widths increase substantially. Above 200~K the lines cannot be resolved as separate lines in the spectra but only via the fitting procedure (see Fig.~\ref{fig:Figure6}). When the resulting linewidths and energies are plotted as a function of temperature, no anomalies can be observed, as shown in Fig.~\ref{fig:Figure7}.

The temperature dependences of all modes are well described by the Klemens model [Fig.~\ref{fig:Figure7}(b)–(d)]. The linewidths increase by almost a factor of three over the investigated temperature range. The resulting phonon–phonon coupling parameters, $\mathrm{\lambda_{ph-ph}}\left(\Alg^{(3)}\right) = 2.8$, $\mathrm{\lambda_{ph-ph}}\left(\Algp\right) = 2.5$, and $\mathrm{\lambda_{ph-ph}}\left(\Algpp\right) = 2.8$, are exceptionally large but consistent with the values found recently for phonon combs in CrSiTe$_3$.\cite{Chen2025} The absolute magnitude of $\mathrm{\lambda_{ph-ph}}$ is not universal and may, in addition to the specific decay channels of the Klemens model, depend on the lattice potential, particularly at high temperature, or even on other decay channels.\cite{PhysRevB.97.054306, Kumar2022, Tiwari2023, Poojitha2022} As a matter of fact, the lines become very narrow already at 80~K, the lowest temperature accessible in our experiment, indicating strongly reduced damping. Simultaneously, the lines in the comb become discernible and gain intensity.

On the basis of the increasing intensity and the equidistance of these lines, we employ a coherent-state formalism as a spectral description of the observed frequency-domain features to our spectra as proposed by Chen and coworkers.\cite{Chen2025} Coherent states proposed first by Schr\"odinger \cite{schrodinger1926coherent} are widely used as a mathematical framework in superconductivity and laser physics to describe the superposition of a large number of oscillators. In the present context, this approach provides an effective description of the frequency components observed in the Raman spectra but does not imply the existence of a time-dependent coherent phonon expectation value in thermal equilibrium. Rather, it implies the simultaneous excitation of several oscillators \cite{Martin:1971} having a very long lifetime. This coherent state is described by an infinite superposition of oscillators,
\begin{equation}
|\langle x \rangle_\omega|^2=(2\pi)^2 e^{-2|\alpha_0|^2}\sum_{n=0}^{\infty} \frac{|\alpha_0|^{4n+2}}{(n!)^2}[\delta(\omega'-\omega+A+An)+\delta(\omega'-\omega+A-An)],
\label{Fourier}
\end{equation}

\noindent where $|\alpha_0|$ represents an effective parameter controlling the relative weight of the frequency components in the coherent-state-based spectral description. $A$ is proportional to the degree of anharmonicity $\lambda^2$ in the Hamiltonian (See Supplementary Material). Eq.~\ref{Fourier} provides the intensity of the frequency components and describes the  distribution of the discrete frequencies $\omega' = \omega - A - An$ and $\omega' = \omega - A + An$. To apply this model to describe our data, we replaced the Dirac delta with a Lorentzian having a finite linewidth (convoluted with a Gaussian for the resolution). The extracted energies and linewidths obtained from both the empirical fit and the applied model were nearly identical to within the error margins. A detailed comparison between the two approaches is presented in Fig.~\ref{fig:Figure8}.
\begin{figure}[t]
  \centering
  \includegraphics[width=85mm]{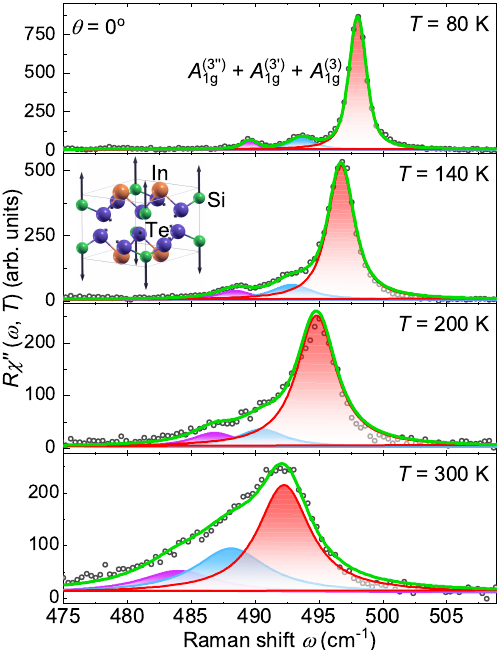}
  \caption{Raman spectra in the range of the $\Alg^{(3)}$ mode at temperatures as indicated. The solid lines represent a Voigt profile fit to the data. All lines become wider with increasing temperature and shift simultaneously to lower energies while maintaining the distance. Inset: Localized vibrations of Te in the SiTe$_3$ tetrahedra associated to the $\Alg^{(3)}$ mode.}
 \label{fig:Figure6}
\end{figure}
\begin{figure}[!t]
  \centering
  \includegraphics[width=85mm]{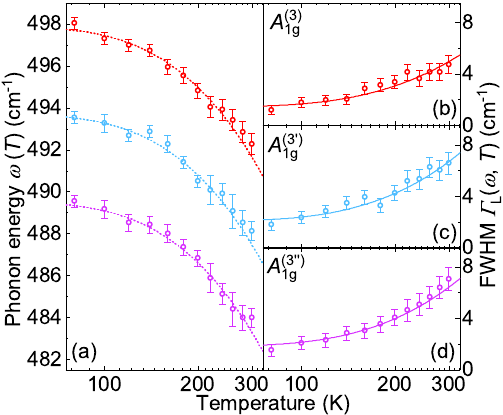}
  \caption{Temperature dependences of energies and linewidths of the $\Alg^{(3)}$ mode and its satellites, $\Algpp$ and $\Algp$ derived from three independent Voigt lineshapes. The equidistant colored dotted lines in (a) represent guide to the eye, with the theoretical difference of 4.2~\wn of neighboring peaks. Solid lines in (b) represent the fit of the linewidth data to the anharmonic model (Eq.~\ref{eq:Anharmonic}).}
 \label{fig:Figure7}
\end{figure}
\begin{figure}[!t]
  \centering
  \includegraphics[width=85mm]{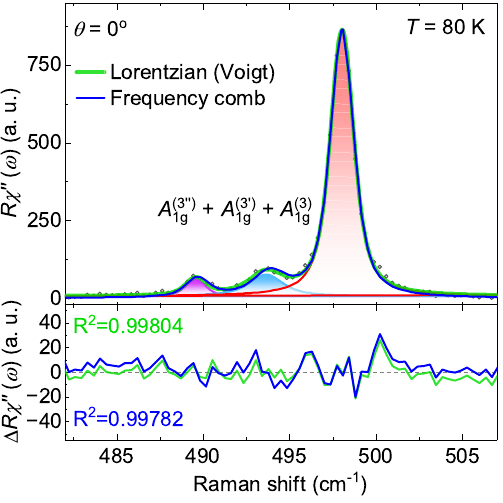}
  \caption{Comparison of coherent-state–based spectral model and individual line model. At 80~K the statistical quality of the frequency comb is only marginally below that of the combination of the three individual lines. (see also Fig.~\ref{fig:Figure6} and Supplementary).}
 \label{fig:Figure8}
\end{figure}
\section*{Discussion}

Coherent optical phonons are typically induced by ultrashort pump pulses. They oscillate in phase and produce distinct signatures in optical and Raman spectra.\cite{ZHAI2024100761} However, equilibrium Raman spectroscopy probes frequency-domain phonon correlations. If the lattice potential has sufficiently strong anharmonic contributions and the phonons are long-lived, coherent-like spectral responses may arise in the absence of ultrafast excitation.\cite{Chen2025}

In the case of \IST, the only phonon expected in the energy range around 500~\wn is a motion of the Si atoms with a negligible involvement of the SiTe$_3$ tetrahedra (Inset of Fig.~\ref{fig:Figure6}). DFT predicts an isolated high-energy Einstein phonon as shown in the Inset of Fig.~\ref{fig:Figure5}. However, instead of a single isolated line a series of equidistant peaks is observed in the Raman spectra (Fig.~\ref{fig:Figure6}), reflecting the underlying nonlinear lattice dynamics. Phonon frequency combs have recently been reported in CrSiTe$_3$ and CrGeTe$_3$.\cite{Chen2025} The comb-like structure was associated with the nonlinear response of a high-energy optical phonon that is spectrally well separated from other phonon states. In \IST, the $\Alg^{(3)}$ mode is separated by a sufficiently large gap in PDOS. Because of this spectral isolation, the mode remains long-lived, as the available decay channels are strongly limited. As a consequence equidistant spectral components emerge. 

An alternative explanation for persistent equidistant peak structure may be thermal "hot-band" progression arising from anharmonic vibrational ladders. For an anharmonic potential in second order perturbation theory we get relatively closely spaced energies at distances $\Delta \omega$.  The intensity of the $n$-th satellite is expected to scale as $I_n/I_0 \propto \exp(-(\hbar \omega - n\hbar\Delta\omega)/k_B T)$, where $\omega = 498.0$~\wn  corresponds to the energy of the $\Alg^{(3)}$ phonon ($n=0$) at 80~\,K, and $\Delta \omega = 4.2$~\wn. As shown in Supplementary Fig.~S1, the experimentally observed temperature dependences of the sattellites at 493.8~\wn ($n=1$) and 489.6~\wn ($n=2$) deviate strongly from this expectation. This discrepancy allows us to exclude a thermal hot-band origin of the observed phonon frequency comb.

In addition, mechanisms based on the thickness or layer dependent quantization can be excluded in the present case. All Raman measurements reported here were performed on bulk single crystals, where the macroscopic sample dimensions prevent the formation of discrete low-energy standing acoustic modes associated with finite-size confinement. In such confined geometries, the characteristic mode spacing is expected to scale with sample thickness or number of layers. No such dependence is observed here.

The satellite peaks observed in \IST are not related to phonon beating either. Phonon beating requires the coexistence of multiple distinct phonon modes with closely spaced frequencies, as demonstrated, for example, for the \Alg modes in Cd$_3$As$_2$ induced by helix vacancies.~\cite{Sun2017} In contrast, for \IST one expects a single isolated $\Alg^{(3)}$ optical phonon branch (Fig.~\ref{fig:Figure6}) with no nearby modes that could give rise to beating. The observed satellites therefore originate from the nonlinear lattice potential leading to additional eigenstates in second order perturbation theory \cite{Chen2025} rather than independent phonon excitations.

In Cr$_2$Ge$_2$Te$_6$ comb-like structures in the phonon spectra were observed recently and interpreted in terms isotope shifts.~\cite{krasucki2025} In \IST, the isotope satellites were to be expected at 9~\wn and 17~\wn below the main line at 500~\wn as opposed to 4.2 and 8.4~\wn here, and their intensities and linewidths exhibit a systematic temperature evolution different from that expected for isotopes.

Finally, we wish to address the dilemma of the strong phonon-phonon coupling of the 500~\wn modes derived via the Klemens model (Eq.~\ref{eq:Anharmonic}), suggesting a rapid decay, and the long lifetime required for phase coherence.  As already pointed out above symmetric decay is not the only mechanism for line broadening. In an anharmonic potential the lines may get broader also by a stronger shift of a the resonance frequencies at elevated temperatures and a superposition of lines at different energies. At low temperatures there are barely any channels for the isolated Einstein mode (and possible satellites) to decay into low-energy phonons as obvious from the DFT phonon dispersion (Inset of Fig.~\ref{fig:Figure5}). The separation from the continuum of other vibrational states by a sufficiently large gap in the PDOS quenches all  decay channels for symmetric decay.  We conclude that the distortion of the harmonic potential required for the formation of the observed frequency comb does not originate from the anharmonic decay of the Einstein-like mode.

The temperature dependences of the fully symmetric low-energy peaks, $A_{1g}^{(1)}$ and $A_{1g}^{(2)}$, change discontinuously in the vicinity of 200~K  (Fig.~\ref{fig:Figure4}). Specifically the line widths do not follow the expected anharmonic variation any longer, indicating a change in phonon–phonon coupling. Along with these discontinuities at 200~K, two-phonon excitations (see Fig.~\ref{fig:Figure5}) become more pronounced. In the Raman spectra, structures beyond the allowed lines may originate either from defect-induced scattering, in which case the spectral features follow a projected PDOS, or from  interactions in the  material.\cite{PhysRevB.97.054306} A comparison of the spectra with the calculated PDOS, in addition to the SEM analysis allow us to rule out the defect scenario. The fact that the two-phonon excitations appear exclusively in the \Alg channel is consistent with the selection rules for two-phonon scattering, where the direct product of two phonon symmetries always contains the totally symmetric representation.\cite{PhysRevB.97.054306} The unusually strong coupling of the \Alg modes in \IST further enhances multi-phonon scattering in this channel, making the overtone features particularly pronounced. As the thermal phonon population and electron occupation increase, these higher-order processes become more probable and the overtones emerge clearly within the PDOS gap. By comparing their positions with the calculated phonon dispersion, the two-phonon features can be linked to low-energy PDOS maxima at roughly half the energies of 2-$ph^1$ and 2-$ph^2$ ($\approx 95$ and $\approx 145$~\wn, respectively), indicating contributions from acoustic or low-lying optical branches (Inset of Fig.~\ref{fig:Figure5}).

These observations reveal that the lattice dynamics of \IST are dominated by strong phonon–phonon interactions and give rise to pronounced anomalies near 200~K. Although a microscopic model for these features is still lacking, we propose that the thermal population of phononic and electronic states in this narrow-gap semiconductor drives these interactions. The appearance of overtone excitations within a PDOS gap supports this interpretation. The formation of a frequency comb, generated by coherent-like phonons and persisting across the full temperature range studied, suggests that the well-isolated Einstein phonon in \IST plays a central role in stabilizing highly structured vibrational spectrum. Together with recent reports of coherent-like phonon states in CrSiTe$_3$ and CrGeTe$_3$,\cite{Chen2025} our findings position Van der Waals trichalcogenides as promising platforms for discovering frequency combs in phonon spectra.

\section*{Methods}

\subsection*{Experimental methods}

Single crystals of \IST were synthesized by melting stoichiometric mixture of In (5N, Alfa Aesar) chunk, Si (5N, Alfa Aesar) lump, and Te (5N, Alfa Aesar) shot. The starting materials were vacuum-sealed in a quartz tube, heated to 1100$^{\circ}$C over 20 hours, held at 1100$^{\circ}$C for 12 hours and then cooled to 700$^{\circ}$C at rate of 1$^{\circ}$C/h.

Scanning electron microscopy was performed using FEI HeliosNanolab 650 scanning electron microscope (SEM) equipped with an Oxford Instruments energy dispersive spectroscopy (EDS) system with an X-max SSD detector operating at 20 kV. Measurements were performed on as-cleaved samples deposited on a graphite tape. To determine the presence of different elements, an area about the size of 100 $\times$ 100 $\mu$m was selected on the deposited material where the EDS analysis was performed.

Inelastic scattering measurements of \IST were conducted using Tri Vista 557 Raman spectrometer in the subtractive backscattering configuration, with the combination of gratings 1800/1800/2400 grooves/mm. As an excitation source Ar$^+$/Kr$^+$ ion laser with 514 \,nm line was used. In this scattering configuration plane of incidence is $ab$ plane ($\vert{a}\vert = \vert{b}\vert$, $\measuredangle(a,b) = $120\grd ) with incident/scattered light propagation direction along crystallographic $c$-axis. The samples were cleaved right before being placed in a high vacuum ($10^{-6}$ mbar), which was achieved with a KONTI CryoVac continuous Helium Flow cryostat with 0.5 mm thick window. Laser beam focusing was accomplished using microscope objective with $\times$50 magnification to a spot size of approximately $8\,\mu{\rm m}$. All Raman spectra were corrected for the Bose factor.

\subsection*{Theoretical methods}

First-principles calculations based on density functional theory (DFT) were carried out using the Quantum ESPRESSO package.\cite{QE-2009} The Perdew–Burke–Ernzerhof revised for solids (PBEsol) exchange-correlation functional \cite{PhysRevLett.77.3865} within the generalized gradient approximation (GGA) was employed to better describe lattice parameters in crystalline solids. The cutoff for wavefunctions and the charge density were set to 50 Ry and 400 Ry, respectively. The $k$-points were sampled using the Monkhorst-Pack scheme, on a 12$\times$12$\times$12 $\Gamma$ - centered grid. For accurate treatment of interlayer interactions, the Van der Waals interactions is included using the Grimme-D2 correction. Phonon frequencies were calculated with the linear response method, as implemented in Phonon package of the Quantum Espresso. The initial atomic structure of \IST was obtained from an experimental CIF file and used as the starting point for first-principles calculations. According to experimental crystallographic data, \IST exhibits trigonal symmetry and belongs to the $P\overline{3}1m$ (No. 162) space group. All calculations correspond to the harmonic approximation at zero temperature and do not include anharmonic or finite-temperature effects.

\section*{Data availability}
Data are available upon reasonable request from the corresponding author.

\section*{Acknowledgements}

Authors are grateful to Vladimir Damljanovi\'{c} for insightful discussions. 

\section*{Funding}

The authors acknowledge funding provided by the Institute of Physics Belgrade, through a grant from the Ministry of Science, Technological Development and Innovation of the Republic of Serbia, Project F-134 of the Serbian Academy of Sciences and Arts. This research was supported by the Science Fund of the Republic of Serbia, 10925, Dynamics of CDW transition in strained quasi-1D systems - DYNAMIQS. This work has received funding from the European Union's Horzion Europe research and innovation programme under grant agreement No 101185375 (HIP-2D-QM). The collaboration between the Institute of Physics Belgrade and the IFW Dresden was supported by the German Academic Exchange Service (DAAD) through project 57703419 and the Ministry of Science, Technological Development and Innovations of the Republic of Serbia. DFT calculations were performed using computational resources at Johannes Kepler University (Linz, Austria). Materials synthesis was supported by the U.S. DOE-BES, Division of Materials Science and Engineering, under Contract DE-SC0012704 (BNL). Electron microscopy was performed at Jozef Stefan Institute, Ljubljana, Slovenia, under Slovenian Research Agency contract P1-0099. C.P. acknowledges financial support from Shanghai Key Laboratory of MFree, China (No. 22dz2260800) and Shanghai Science and Technology Committee, China (No. 22JC1410300).

\section*{Author contributions statement}

A.M., S. Dj. M., T.B., J.B., R.H. and N.L. conducted the Raman experiments and analyzed the results. B.V. performed SEM measurements and analyzed the results. A.\v{S}. and J.P. performed theoretical calculations. Y.L. and C.P. synthesized the sample and analyzed the results. E.S.B. and Z.V.P. analyzed the results.
All authors were included in the writing and review of the manuscript.

\section*{Competing interests}
The authors declare no competing interests.

\end{document}